\documentclass[12pt, draftclsnofoot, onecolumn]{IEEEtran}
\pdfoutput=1

 \usepackage{cite,url}
 \usepackage{amsmath,amsfonts,amssymb,verbatim}
 \usepackage{amsmath,amssymb,graphicx,bm,color,amsbsy,amsthm,algorithm,algorithmic,epsfig,subfigure}
 \usepackage{epstopdf}
 \usepackage{subfiles,caption}
 \usepackage[utf8]{inputenc}
 \usepackage[english]{babel}
 
 \newtheorem{theorem}{Theorem}

\DeclareMathOperator*{\maximize}{maximize}
\DeclareMathOperator{\subjectto}{subject~to}

\begin{document}

\title{
	Joint Routing and Resource Allocation for Millimeter Wave Picocellular Backhaul}
\author{\IEEEauthorblockN{ Maryam Eslami Rasekh\IEEEauthorrefmark{1}, Dongning Guo\IEEEauthorrefmark{2}, Upamanyu Madhow\IEEEauthorrefmark{3}}
\IEEEauthorblockA{Department of Electrical and Computer Engineering, 
University of California Santa Barbara\\
Email:\{\IEEEauthorrefmark{1}rasekh, \IEEEauthorrefmark{3}madhow\}@ece.ucsb.edu}
\IEEEauthorblockA{Department of Electrical Engineering and Computer Science, 
Northwestern University, Evanston, IL\\
Email: \IEEEauthorrefmark{2}dGuo@northwestern.edu}
}

\maketitle
	
\begin{abstract}
Picocellular architectures are essential for providing the spatial reuse required to satisfy the ever-increasing demand for mobile data. A key deployment challenge is to provide backhaul connections with sufficiently high data rate. Providing wired support (e.g., using optical fiber) to pico base stations deployed opportunistically on lampposts and rooftops is impractical, hence wireless backhaul becomes an attractive approach. A multihop mesh network comprised of directional millimeter wave links is considered here for this purpose. The backhaul design problem is formulated as one of joint routing and resource allocation, accounting for mutual interference across simultaneously active links. A computationally tractable formulation is developed by leveraging the localized nature of interference and the provable existence of a sparse optimal allocation. Numerical results are provided for millimeter (mm) wave mesh networks, which are well suited for scaling backhaul data rates due to abundance of spectrum, and the ability to form highly directional, electronically steerable beams.
\end{abstract}

\begin{IEEEkeywords}
		Millimeter Wave, 5G, wireless backhaul, mesh network, resource allocation, medium access control, routing, interference.
\end{IEEEkeywords}

\section{Introduction}

The wireless industry is striving
to keep up with mobile data demand from smart devices and data-hungry applications. 
Picocellular architectures comprised of closely-spaced access points with intense spatial reuse play a critical role in the evolution of mobile systems, particularly in high-density urban and suburban environments \cite{ge20145g}. LTE data rates are projected to approach Gigabits per second peak rates through carrier aggregation, which is likely to be extended further in next generation networks by using 60 GHz - or other unlicensed millimeter (mm) wave bands - directly from pico base station to the mobile \cite{Zhu_Mobicom2014}, 
assuming that significant challenges due to blockage and mobility can be overcome.  Indeed, a recent interference analysis for such networks  \cite{marzi2015interference} indicates that capacities of the order of 
terabits per second per kilometer (along a single urban canyon)
can be obtained with only a few GHz of spectrum by taking advantage of the aggressive spatial reuse enabled by highly directional mm wave links. Moreover, this capacity roughly adds up across parallel canyons, given the strong isolation provided by building blockage.

Delivering such high data rates to mobile users requires that the pico base stations have a sufficiently high-capacity backhaul connection to the Internet. For opportunistic picocellular deployments on lampposts and rooftops, providing optical fiber connectivity for each base station is practically impossible in the foreseeable future, so that
wireless backhaul becomes a natural choice~\cite{
	singh2015tractable, 
	zheng201510gbps, 
	taori2015point-to-multipoint}.
In this paper, we consider a mesh network with highly directive mm wave links as a means of extending backhaul from wired gateways to the picocell access points. Each access point is a node in the wireless mesh network and is connected to neighboring nodes through high-speed directional mm wave links.  The objective is to route traffic between base stations and gateways through multihop paths,
such that each picocell can support a given level of downlink and uplink throughput on the access links to mobile devices.  We assume that directional antennas are
used on each backhaul link, but because of half duplex communication and residual interference, these links cannot be treated as wires. In particular, in long urban canyons where even non-contiguous links are likely to be aligned, the interference between distant links must be taken into account in deriving the optimal resource allocation and routing.

\subsection{Approach and contributions}

We formulate the wireless mesh backhaul design problem as a joint routing and 
resource allocation optimization, with the goal of maximizing the access rate at the base stations, while accounting for the mutual
interference between simultaneously active links.  Our framework applies to any mesh backhaul network via the following abstraction:
For any set of simultaneously active links, we must be able to compute the achievable data rate on each active link while accounting for the
interference from other active links.  These interference patterns, of course, depend on
the propagation environment, the antenna patterns, and the carrier frequency.
For example,  in built-up urban environments,  
the highly directive nature of the links, and the ease of blockage of mm waves by buildings, 
imply that each link incurs interference only with links within the same street.  However, the interference among such links can be significant 
due to their similar alignment. In more open suburban environments, on the other hand, a larger number of links can cause interference on each other, but links are less likely to be aligned, so that interference is typically severely attenuated by antenna directionality at the transmitter and the receiver.

Key novel contributions of the proposed architecture and analysis include the following:

\begin{itemize}
	\item {\textit{Accurate interference modeling:} Environment-specific propagation models are used to derive a realistic interference graph for the network and the optimal level of spatial reuse is maintained, yielding the highest possible backhaul throughput upon deployment. The solutions proposed here can easily be generalized to a variety of network structures with different antenna patterns and arbitrary interference models.
	}
	
	\item{\textit{Arbitrary traffic flow:} We first develop a framework for \textit{downlink} traffic routing and resource allocation. We then extend the formulation to perform joint uplink and downlink optimization. 
	}
	
	\item{\textit{Arbitrary topologies:} The proposed optimization framework applies to arbitrary network structures with limited node degrees. The networks in our simulations are comprised of several gateways that are connected to all nodes through multihop paths. Each access point is connected to all neighbors that are close enough to form a connection (the number of neighbors is not large in practice). The existence of various paths from each pico base station to different gateways provides redundancy in network resources and improves backhaul reliability, allowing the network to adapt to disruption of links due to blockage or hardware failure. 
	}
	
	\item{\textit{Versatile and efficient optimization framework:} The problem of resource allocation for optimizing backhaul throughput is first formulated as a linear program, the dimensions of which grow exponentially with network size. Demonstrating the existence of a sparse solution to this linear program, we formulate an equivalent mixed-integer linear program that scales linearly with network size. The proposed formulation is able to solve relatively large networks with hundreds of nodes in a short period of time.}
	
\end{itemize}

\subsection{Related work}

This paper consolidates and significantly extends the work in our prior conference publication \cite{rasekh2015interference}, where we present a combinatorial optimization framework for optimal resource allocation and routing.  In \cite{rasekh2015interference}, resources are divided between all possible activation patterns, or {\it subsets} of simultaneously active links. This blows up the problem size exponentially, going from a scheduling of $L$ links to $2^L$ possible link combinations, but enables formulation of backhaul scheduling as a linear program. Solving this problem is straightforward for small networks but quickly becomes intractable for larger graphs, and requires suboptimal clustering of the larger network. One important result, guaranteed by Caratheodory's theorem \cite{meurant2014handbook}, is that a sparse solution exists for the combinatorial formulation with at most $N$ active patterns, where $N$ is the number of nodes in the network. In this paper, we first revisit the problem in  \cite{rasekh2015interference} and provide a scalable optimization framework which exploits the existence
of an $N$-sparse solution. We reformulate the problem in terms of {\it local} interference parameters. 
The result is a binary linear program (BLP) that scales near-linearly with network size, and can be solved relatively quickly using branch and bound techniques for larger networks with multiple gateways and hundreds of nodes. We also incorporate joint scheduling for uplink and downlink traffic, while the analysis of \cite{rasekh2015interference} only optimizes for downlink support.

The core idea of considering all possible link activation patterns 
is inspired by the approach used in \cite{zhuang2015traffic-driven,zhou20171000,kuang2016optimal,zhuang2015energy-efficient} for 
optimizing downlink spectrum allocation in cellular networks.
Following \cite{zhuang2017scalable}, we reformulate the convex combinatorial resource allocation problem into a scalable mixed integer optimization problem. The problem considered here is fundamentally different from that of \cite{zhuang2015traffic-driven,zhou20171000,kuang2016optimal,zhuang2015energy-efficient,zhuang2017scalable} due to the multihop nature of the network and the added problem of routing both uplink and downlink traffic.

Increasing cellular capacity through self-backhauled small cells has been the subject of many previous works. Some studies consider placement of relay nodes inside cells to improve signal quality at cell edges \cite{li2015small,peters2009future}, yet the capacity boost obtained through simple radio frequency (RF) amplification and relaying is limited. Other studies consider addition of small-cell base stations inside a macrocell, each receiving wireless backhaul directly from the wired macro-BS, forming a backhaul network with star topology \cite{singh2015tractable,li2017joint,karamad2015optimizing}. In \cite{du2017gbps}, a multihop network is considered in the form of a row of nodes inside one street along with one wired node that acts as a gateway. These simplified structures are useful to provide insight into the capacity of wirelessly backhauled picocells, but evaluating and optimizing a realistic network requires increasing the scope to include multihop paths along different streets and networks with several gateways. 

Several papers consider the general problem of multihop mesh backhauling, but limit the number of links on each node to a single steerable beam \cite{yuan2017optimal,singh2015tractable,garcia2015analysis}. Such a model has two basic limitations; first, limiting nodes to a single RF chain does not take advantage of the potential for spatial reuse provided by highly directional mm wave arrays that allow simultaneous transmission (or reception) on multiple links at the same node. Second, the beam steering capacity of mm wave phased arrays is over-estimated by this approach. Planar arrays can at most beamform within a 180-degree portion of azimuth, even less if element pattern imperfections are taken into account. To cover the full angular domain, a node will require at least three or four planar faces. While it is possible to control several array faces with a single RF chain, most standard front-ends are not designed in this manner. 

In modeling the interference behavior of links, most studies assume a binary effect: any two links either collide completely (mostly by contradicting the half duplex constraint) and have to be orthogonalized in resources, or have zero interference \cite{yuan2017optimal,narlikar2010designing}. 
Few papers take note of the residual interference in the network and include it in the allocation optimization. In \cite{garcia2015analysis} a general multihop mesh network has been modeled with a detailed interference model incorporated in the optimization. However, the mesh nodes are assumed to have a single RF chain allowing for at most one link connected to each node being active simultaneously. In \cite{kulkarni2018many}, a multihop backhaul network was modeled as a uniform square grid of nodes to provide analytical insight into throughput capacity as a function of the size of the cluster supported by each gateway. An interesting observation was that backhaul capacity is not diminished as a result of residual interference, and with careful scheduling of interfering links the interference-free capacity of the network can be achieved. We demonstrate the importance of interference avoidance in scheduling and note that the proposed framework provides such a schedule for their specific topology.

\subsection{Outline}

The remainder of this paper is organized as follows. The system model is described in Section \ref{sec:system_model}. In Section \ref{sec:combinatorial_formulation} we formulate the routing and scheduling problem and prove the existence of a sparse solution. We then describe the construction of a scalable formulation on this basis in Section \ref{sec:scalable_formulation}. We present and discuss simulation results in Section \ref{sec:results}. Conclusions and possible directions for future work are discussed in Section \ref{sec:conclusions}.

\section{System Model \label{sec:system_model}}

\begin{figure}
	\centering
	\subfigure[]{
		\includegraphics[width=.60\columnwidth]{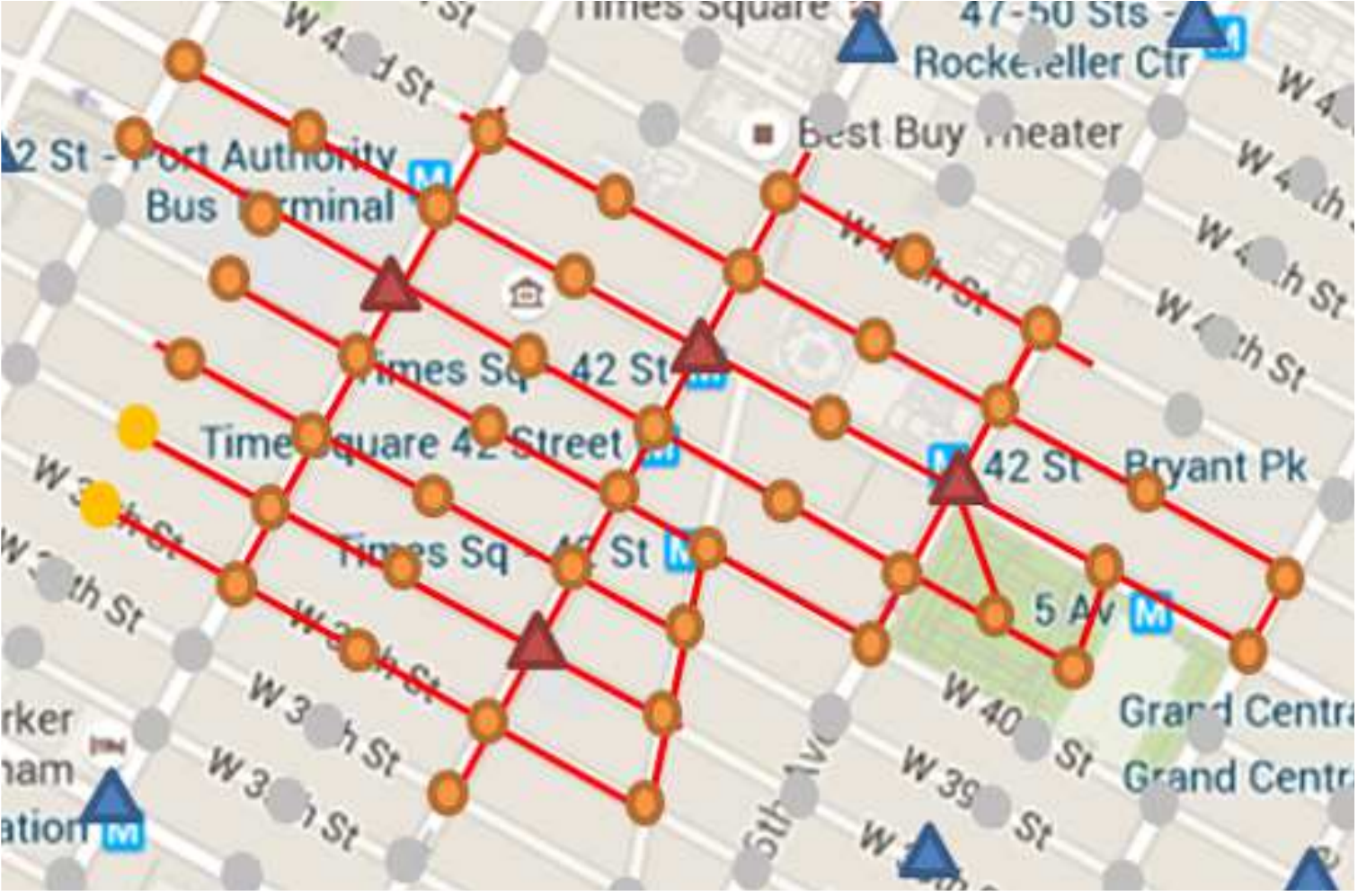} \label{fig:topology_urban}
	}
	\subfigure[]{
		\includegraphics[width=.73\columnwidth]{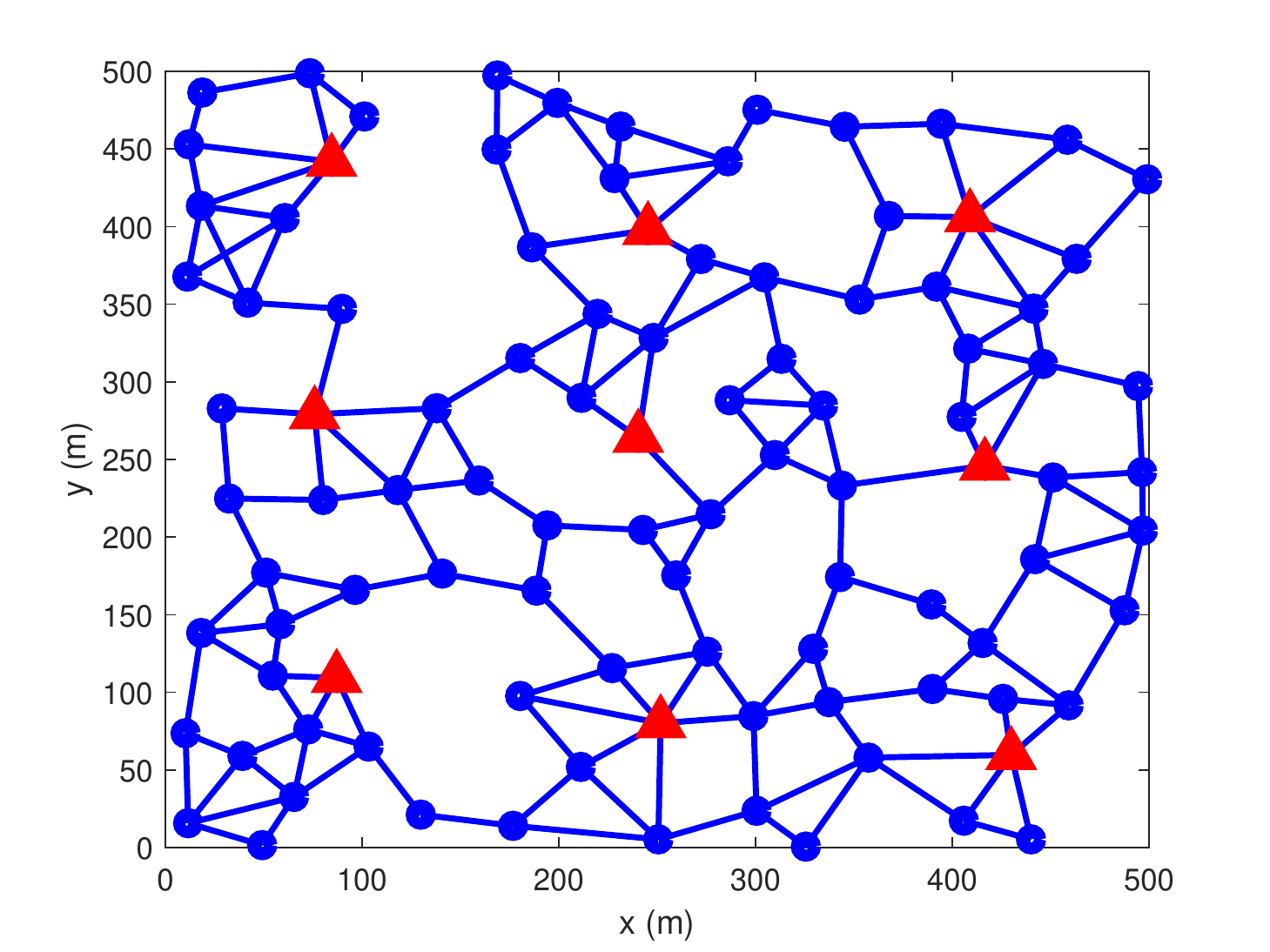} \label{fig:topology_rural}
	}
	\caption{Example of (a) a 48-node, 4-gateway portion of an imagined picocellular backhaul network in Manhattan, (b) a randomly generated network of  100 nodes and 9 gateways in a 500 m$\times$500 m area, using the suburban propagation model. } \label{fig:topology}
\end{figure}

We envision an outdoor picocellular network with cell radii as small as tens of meters. The pico base stations are placed opportunistically on existing structures such as lampposts, building walls, and ceilings.
Each base station, 
or ``node", 
is connected to several nearby nodes (neighbors) through directional mm-wave links. 
Only a fraction of nodes receive wired backhaul connection to the communication infrastructure
. 
These nodes operate as gateways and the downlink and uplink data of each node is routed through a multihop path in the mesh to one of the gateway nodes. Examples of this structure are shown in Fig. \ref{fig:topology} where the gateways are marked by triangles and non-wired base stations by circles. Each line connecting two neighboring nodes represents two wireless links, one in each direction. Phased array antennas are utilized at either end of the link to maintain directional transmit and receive radiation patterns and reduce interference. We define the nominal link SNR and nominal link rate as the SNR and throughput of links in the absence of interference, and transmit power is controlled such that these values are the same for all links.

We denote by $\Gamma =\{1,...,N\}$ the index set of all $N$ non-gateway nodes in the network 
and by $\Lambda = \{1,...,L\}$ the index set of all links in the network. 
As far as interference allows, any number of links connected to a node can be active simultaneously.
However, because a transmitted signal is generally strong enough to saturate all co-located receivers, communication is half-duplex, namely, 
the links connected to one node can transmit simultaneously, or receive simultaneously, but at no time do some links transmit and some receive.
As the opposing directions between two nodes are regarded as two separate links in the proposed optimization framework, the duplexing constraints can be easily incorporated by assuming infinite (or disabling) interference levels for such conflicting links.

To find the optimal allocation for interference management and routing, we  apply a combinatorial approach  where the available resources (e.g., time and/or frequency) are divided between different {\it subsets} of links with the objective of maximizing the minimum backhaul throughput delivered to all nodes in the network. 
A general scheduling framework can be defined as follows. The total time in each frame is divided into fragments of variable lengths and during each fragment (or slot) a certain combination of links are active. We denote each possible partitioning of links into on and off by an ``activation pattern", which determines the amount of interference each link is subject to, and hence its data rate. The resource management problem is equivalent to 
allocating 
to every possible activation pattern an appropriate fraction of all resources (which may be none). 
For every pattern $A\subset\Lambda$, let $x_A\in[0,1]$ denote the fraction of resources allocated to that pattern, which must satisfy the resource constraint, 
\[ \sum_{A\subset \Lambda} x_A =1. \] 
The problem of allocating resources to $L$ links becomes that of allocating non-overlapping portions to the $2^L-1$ non-empty subsets of links. 

We consider two distinct deployment scenarios, namely the suburban and urban setting
. In the suburban setting, antennas are placed on rooftops that are of relatively similar heights. As a result, the street geometry does not have a significant effect on the channel as links are assumed to be line-of-sight (LOS) without any blocking structures between antennas. The free-space propagation model is thus used to model the channel between different nodes and the only factors determining signal and interference strength are the radiation pattern of antennas and distances between nodes. In the high-rise urban setting, antennas are mounted on below-rooftop-level structures such as lamp-posts, traffic lights, and building walls. In this case a street-canyon model is considered wherein the channel is a combination of the LOS path and single-bounce reflections from the two canyon walls and ground. Transmit and receive radiation patterns and Fresnel (specular) reflection loss are accounted for in the model, as well as phase offset between propagation paths that is assumed to be uniformly random over the span of $[0,2\pi]$ radians. The model parameters used in this scenario are reported in Table \ref{tab:parameters}.

\begin{table}
	\centering
	\caption{Simulation parameters}
	\renewcommand{\arraystretch}{1.5}
	\begin{tabular}{| l | c |}
		\hline
		nominal link SNR & 10 dB \\ \hline
		array size & 32 \\ \hline
		street width & 25 m \\ \hline
		distance of node from wall & $\sim U(4,21)$m \\ \hline
		antenna height & $\sim U(5,8)$m \\ \hline
	\end{tabular}
	\label{tab:parameters}
\end{table}

In simulations, we consider the topology of Fig. \ref{fig:topology_urban} for the urban scenario, and randomly generated topologies similar to Fig. \ref{fig:topology_rural} for the suburban scenario. These are generated by placing nodes uniformly at random in a square area with a minimum link distance of 10 m. For each node, a random number is drawn between 3 and 5, then the node is connected to that number of its nearest neighbors, and links that are longer than a threshold are truncated. This threshold is tuned to obtain a well connected graph and is set high enough to ensure that no node is left disconnected from the network. The maximum node degree is set to 6 for gateway nodes since that would be an appropriate design practice to increase the backhaul capacity of the network.
The number of gateways is chosen to be approximately 10\% of the total number of nodes, and gateway nodes are defined by choosing the closest nodes to uniformly placed anchor points in the area. While this generating scheme obtains relatively realistic topologies, the occurrence of sub-par configurations that impose bottlenecks to service of some areas is possible. In practice, node placements would be optimized to some extent to improve the capacity of the network and better performance than the simulated outcomes can be expected. 

\section{Combinatorial Formulation of the Allocation Problem \label{sec:combinatorial_formulation}}

In this section we formulate an optimization problem that maximizes the minimum backhaul throughput delivered to every node subject to interference and resource constraints. 
When pattern $A\subset \Lambda$ is active, the throughput of link $l\in A$ is equal to,
\begin{align}
\gamma_{l,A} & =
\log{\left ( 1+ \frac{S_l}{n+\sum_{k\in A\backslash \{l\}}I_{k\rightarrow l}}\right )} 
\label{eq:link_rate}
\end{align}
where $S_l$ is the signal power spectral density (PSD) of link $l$, $I_{k\rightarrow l}$ is the interference PSD from link $k$ on link $l$, and $n$ is the noise PSD. While $\gamma_{l,A}$ is left undefined if $l\notin A$ to avoid redundant parameters, it is equivalent (and natural) to think of it as equal to 0 if $l\notin A$. The total data rate of link $l$, i.e., data transferred in a unit of time, is therefore obtained by,
\[r_l=\sum_{A\subset \Lambda : l\in A}x_A \gamma_{l,A}.\]

\subsection{Optimization of downlink only}
In the downlink, each node is considered as a sink of its own traffic. The gateway nodes are sources with no throughput constraint. Although this is a network of multiple ``commodities", each intended for one non-gateway node and potentially delivered via a combination of multiple flows, it is equivalent to a network of a single commodity by using the following insight: We add a virtual source node where all flows originate with an unlimited link to each of the gateway nodes. 
We also add an unlimited link from every non-gateway node to a virtual sink node where all flows end. This allows us to combine all traffic as a single commodity from the virtual source to the virtual sink. It then suffices 
to formulate the routing problem by imposing constraints on nodes and links instead of flows. 

The downlink data rate delivered to node $i$, normalized to system bandwidth, is thus equal to,
\begin{equation}
d_i = \sum_{l\in I_i} r_l - \sum_{k\in O_i} r_k \,\, , \quad  \quad i\in \Gamma
\label{eq:flow_constraint}
\end{equation}
where $I_i$ is the set of links that flow into node $i$ \footnote{Not to be confused with $I_{k\rightarrow l}$ which denotes the interference of link $k$ on link $l$.} and $O_i$ is the set of links that flow out of node $i$. 

If allocations were in the frequency domain, the spectral efficiency parameters $\gamma_{l,A}$ would depend on the allocation variables $x_A$ 
as the link PSD may depend on the bandwidth. Time domain allocation thus simplifies the formulation.

The network utility is in general a function of the rate vector $[r_1,...,r_L ]^T$. In this paper,
the optimal  allocation maximizes the minimum node service $d$ subject to the resource constraint. This allocation can be obtained by solving the following optimization problem: 

\begin{subequations}
	\label{eq:basic_formulation}
	\begin{align}
	\maximize_{[x_A],[r_l],[d_i], d} \quad
	& \quad d  	\label{eq:basic_formulation_objective}\\
	\subjectto \quad
	& \sum_{A\subset\Lambda} x_A = 1, \label{eq:basic_formulation_resource}\\
	& r_l = \sum_{A\subset\Lambda:l\in A} x_A \gamma_{l,A}  , & l\in \Lambda \label{eq:basic_formulation_link_rate}\\
	&
	\sum_{l\in I_i}r_l - \sum_{k\in O_i}r_k = d_i , \, & i\in \Gamma
	\label{eq:basic_formulation_flow}\\  
	&  d_i \geq \alpha_i d, \, & i\in \Gamma 
	\\
	& x_A \ge 0, & A\subset \Lambda
	\end{align}  
\end{subequations}
where $r_l$ is the data rate on link $l$, and $\alpha_i$ is a weighting factor that can be used to facilitate uneven service that provides larger backhaul throughput for high-traffic hotspots if needed. In our simulations we assume uniform traffic at all nodes and set $\alpha_i=1, \forall i$.
Note that the node flow constraints 
only apply to non-gateway nodes. 

Both the objective and constraints in the optimization 
are linear, therefore an optimal allocation is found by solving a linear program with around $2^L$ variables. Of course, if all of these possible patterns receive non-zero time allocation, the complexity of scheduling would grow exponentially with network size. However, there is always a sparse optimal solution to (\ref{eq:basic_formulation}) in 
the following sense:

\begin{theorem}
	For a network of $L$ links, $N$ non-gateway nodes, and one or more gateways, there exists a solution $[x_A]_{A\subset \Lambda}$ to the downlink scheduling problem of (\ref{eq:basic_formulation}) that is at most $N$-sparse. 
	\label{theorem:caratheodory}
\end{theorem}
{\it \noindent Proof:} Consider the $N$ dimensional vector ${\bf d}=[d_1,...,d_{N}]^T$ that is a feasible  point for (\ref{eq:basic_formulation}) achieved by the allocation $[x_A]$. The maximum utility is determined by ${\bf d}$, which can be written as a convex combination of $2^L-1$  vectors of dimension $N$ as,
\[
\begin{bmatrix}
d_1 \\ d_2 \\ \vdots \\ d_{N}
\end{bmatrix}
= \sum_{A\subset \Lambda} x_A
\begin{bmatrix}
\sum_{l\in I_1\cap A}  \gamma_{l,A} - \sum_{k\in O_1\cap A}  \gamma_{k,A}\\
\sum_{l\in I_2 \cap A}  \gamma_{l,A} - \sum_{k\in O_2 \cap A}  \gamma_{k,A} \\
\vdots \\
\sum_{l\in I_N \cap A}  \gamma_{l,A} - \sum_{k\in O_N \cap A}  \gamma_{k,A}
\end{bmatrix}.
\]
Hence  ${\bf d}$ resides in a polyhedron in $\mathbb{R}^N$ with up to $2^L-1$ corner points. 
Noting the resource constraint, $\sum_{A\subset \Lambda}x_A=1$, Caratheodory's theorem \cite{meurant2014handbook
} states that ${\bf d}$ can be written as a convex combination of at most $N+1$ of these $2^L-1$ points. Therefore there exists an allocation $[x'_A]$ satisfying $\sum_{A\subset\Lambda}x'_A=1$ with at most $N+1$ nonzero entries that yields the same node service vector as $[x_A]$, i.e.,
\begin{align*}
&d_i= \sum_{A\subset\Lambda} x'_A \left( \sum_{l\in I_i \cap A} \gamma_{l,A} - \sum_{k\in I_i \cap A} \gamma_{k,A} \right). \qquad i\in \Gamma
\end{align*}
An optimal ${\bf d}$ cannot be an interior point of the polyhedron; otherwise one may strictly increase all of its elements to a boundary point, which increases the objective $d$. This implies that there exists an $N$-sparse optimal allocation,
hence the proof of Theorem \ref{theorem:caratheodory}. $\qed$

In the networks of interest to us, the interference matrix can be very sparse, resulting in many possible optimal allocations when we solve the linear program. 
By adding very small random fluctuations to the interference matrix, the solution can be made unique with high probability, and since a sparse solution is guaranteed to exist, we will find one using this trick.

One problem that arises when solving (\ref{eq:basic_formulation}) is that the objective function does not penalize suboptimal routing as long as the delivered throughput is unchanged. As a result, a shorter path toward a node is not differentiated from a longer path with more hops, which can result in unnecessarily long paths and excessive latency and power consumption. This can be prevented by adding a linear term to the objective that implicitly penalizes delay. Assuming transfer of data on each link represents one unit of delay in the network, the sum of all link data rates can be taken as a linear proxy for delay and power consumption. This is done by changing the objective of   (\ref{eq:basic_formulation_objective}) to
\begin{equation}
\label{eq:global_formulation}
d - \lambda \sum_{l\in\Lambda} r_l,
\end{equation} 
where the weighting factor $\lambda$ is chosen to be small enough to ensure service rate, $d$, is always prioritized over the delay penalty term. It is shown in \cite{rasekh2015interference} that a sufficient condition to enforce this priority is 
\begin{equation}
\label{eq:priority}
\lambda < \frac{1}{L\sum_i
	\alpha_i}
\end{equation}
which, in the case of uniform service to all nodes ($\alpha_i \equiv 1$), simplifies to 
\[ \lambda < \frac{1}{LN}. \]

\subsection{Joint uplink-downlink optimization}
In this section, we extend the formulation of (\ref{eq:basic_formulation}) to include both uplink and downlink. While the single commodity model no longer applies, we can add an additional pair of virtual source and virtual sink nodes for uplink traffic, which yields a network of two commodities. We formulate the optimization by defining two sets of uplink and downlink service rates, $[d_i]$ and $[u_i]$, and two sets of link rate variables, $[r^\text{d}_l]$ and $[r^\text{u}_l]$, that satisfy
\begin{align}
&\sum_{l\in I_i} r^\text{d}_l  - \sum_{k\in O_i} r^\text{d}_k = d_i ,   \nonumber \\ 
& \sum_{l\in O_i} r^\text{u}_l - \sum_{k\in I_i} r^\text{u}_k = u_i .  \nonumber
\end{align}
The rate on each link will be the sum of the rate supporting downlink and uplink data, and (\ref{eq:basic_formulation_link_rate}) will be modified to
\[ r^\text{d}_l + r^\text{u}_l = \sum_{A\subset \Lambda:l\in A}x_A \gamma_{l,A}.
\]

The optimal allocation is hence obtained from solving the following optimization problem, wherein the constants $\alpha_i$ and $\beta_i$ determine the relative downlink and uplink traffic of different nodes.

\begin{subequations}
	\label{eq:basic_formulation_ud}
	\begin{align}
	&\maximize_{[x_A],[r_l^\text{d}],[r_l^\text{u}],[d_i],[u_i],c} \quad
	\quad  c 	\label{eq:basic_formulation_objective_ud}\\
	&\subjectto \nonumber \\
	& \quad \sum_{A\subset\Lambda} x_A \le 1 \label{eq:basic_formulation_resource_ud}\\
	& \quad r^\text{d}_l + r^\text{u}_l = \sum_{A\subset \Lambda:l\in A} x_A \gamma_{l,A}, & l \in \Lambda \\
	& \quad \sum_{l\in I_i} r^\text{d}_l  - \sum_{k\in O_i} r^\text{d}_k = d_i , \, & i\in \Gamma \label{eq:basic_formulation_flow_d}\\  
	& \quad \sum_{l\in O_i} r^\text{u}_l  - \sum_{k\in I_i} r^\text{u}_k = u_i , \, & i\in \Gamma \label{eq:basic_formulation_flow_u}\\  
	& \quad d_i \geq \alpha_i c \, , \quad u_i \geq \beta_i c, \, & i\in \Gamma \nonumber 
	\\
	& \quad r^d_l \ge 0 \, . \qquad  r^u_l \ge 0 & l\in \Lambda \nonumber
	\end{align}  
\end{subequations}

For this formulation, Caratheodory's theorem cannot be applied to the node rate inequalities as easily as in Theorem \ref{theorem:caratheodory}, however a similar argument can be made for the link rate vector $[r^\text{d}_1+r^\text{u}_1,...,r^\text{d}_L+r^\text{u}_L]$ that is a convex combination of the $2^L-1$ points (enumerated by $A$) in $L$ dimensional space,  $[\gamma_{1,A},...,\gamma_{L,A}]^T$. For any allocation $\textbf{x}=[x_A]_{A\subset\Lambda} $ that is able to support rate vectors $\textbf{r}^\text{d}$ and $\textbf{r}^\text{u}$ and, equivalently, node downlink and uplink vectors $\textbf{d}$ and $\textbf{u}$, there exists an $L$-sparse allocation $\textbf{x}'$ that provides the exact same link rates and uplink and downlink node service rates. Thus a solution with no more than  $L$ active patterns can be guaranteed to exist for any target rate vector pair.

The formulations presented in this section are suitable for relatively small networks. As $L$ increases to even moderate sized values, the number of variables in the problem grows exponentially until the time or space (memory) complexity becomes unmanageable.
In the next section, a  scalable reformulation is developed.

\section{A Scalable Reformulation \label{sec:scalable_formulation}}
One characteristic of the network that can be leveraged to reduce the problem size is the localized nature of interference, which allows decoupling of constraints between distant areas of the network. 
We first define the neighborhood of link $l$ as the set of links, $\Lambda_l$,  that cause non-negligible interference on it, i.e., whose signal strength at the receiving end of $l$ is above a  threshold. 
Subsequently, the ``local patterns" of link $l$ are all possible subsets of its neighborhood, $B\subset \Lambda_l$. A local pattern $B$ is in effect when all links in $B$ are active {\it and} all links in $\Lambda_l\backslash B$ are inactive. By this definition, the spectral efficiency of a link only depends on the activation pattern of links in its neighborhood, or its {\it local} activation pattern.
To maintain a pessimistic estimate of throughput, the interference of all links outside the neighborhood are added to the noise and interference level when calculating throughput. In reality, many of the links outside the neighborhood will not be active, but we can ensure this worst-case assumption does not affect the result significantly by setting the threshold to be low enough. 
There is thus a trade-off between computational complexity and accuracy: Choosing a low threshold yields more exact results at the expense of increasing neighborhood size and enlarging the problem. In our simulations, we set the interference threshold to 3 dB below the noise level, assuming a nominal SNR of 10 dB. We find the disparity between the pessimistic and actual throughput to be less than 1\% in all simulated cases.

Similar to (\ref{eq:basic_formulation}), we define the {\it local} allocation variable $x_l^B$ that is the resource allocated to local pattern $B\subset \Lambda_l$ of link $l$. When enumerating local patterns, the empty set is also counted since a nonempty global pattern may activate none of the links in one neighborhood. The data rate on link $l$ is thus equal to
\[
r_l = \sum_{B\subset \Lambda_l:l\in B} x_l^{B} \gamma_l^B , \quad l\in\Lambda
\]
where  $\gamma_l^B$ is the spectral efficiency of link $l$ under local pattern $B$, derived (pessimistically) as
\begin{align*}
\gamma_l^B &= \log \left( 1+\frac{S_l}{n+\sum_{k\in B\backslash \{l\}} I_{k\rightarrow l} +\sum_{j\notin \Lambda_l} I_{j\rightarrow l}  } \right).
\end{align*} 

Recall that Theorem \ref{theorem:caratheodory} ensures the existence of an optimal solution to (\ref{eq:basic_formulation}) that activates at most $N$ patterns. We therefore consider a segmentation of the unit time frame to $N$ slots, indexed by $M=\{1,...,N\}$. The global slots are of variable lengths, denoted by $\{ y_m \}_{ m\in M}$, that satisfy the resource constrains, $\sum_{m\in M} y_m = 1$. 
Let $P_m$ denote the {\it global} pattern activated in the $m$-th slot. Then $P_m \cap \Lambda_l$ is the corresponding \textit{local} pattern in the neighborhood of link $l$ in the $m$-th slot. These local patterns may overlap and their union is $P_m$.
We define the augmented local allocation  variables $x_l^{B,m}$ as the resource allocated to local pattern $B$ of link $l$ in slot $m$. Evidently, for every $l\in \Lambda$, $m\in M$, and $B\subset \Lambda_l$,
\begin{align*}
& x_l^{B,m}  = \begin{cases}
y_m & \text{if } B=P_m\cap \Lambda_l, \\
0 & \text{otherwise.}  
\end{cases}
\end{align*}

\begin{figure}
	\centering
	\includegraphics[width=.75\columnwidth]{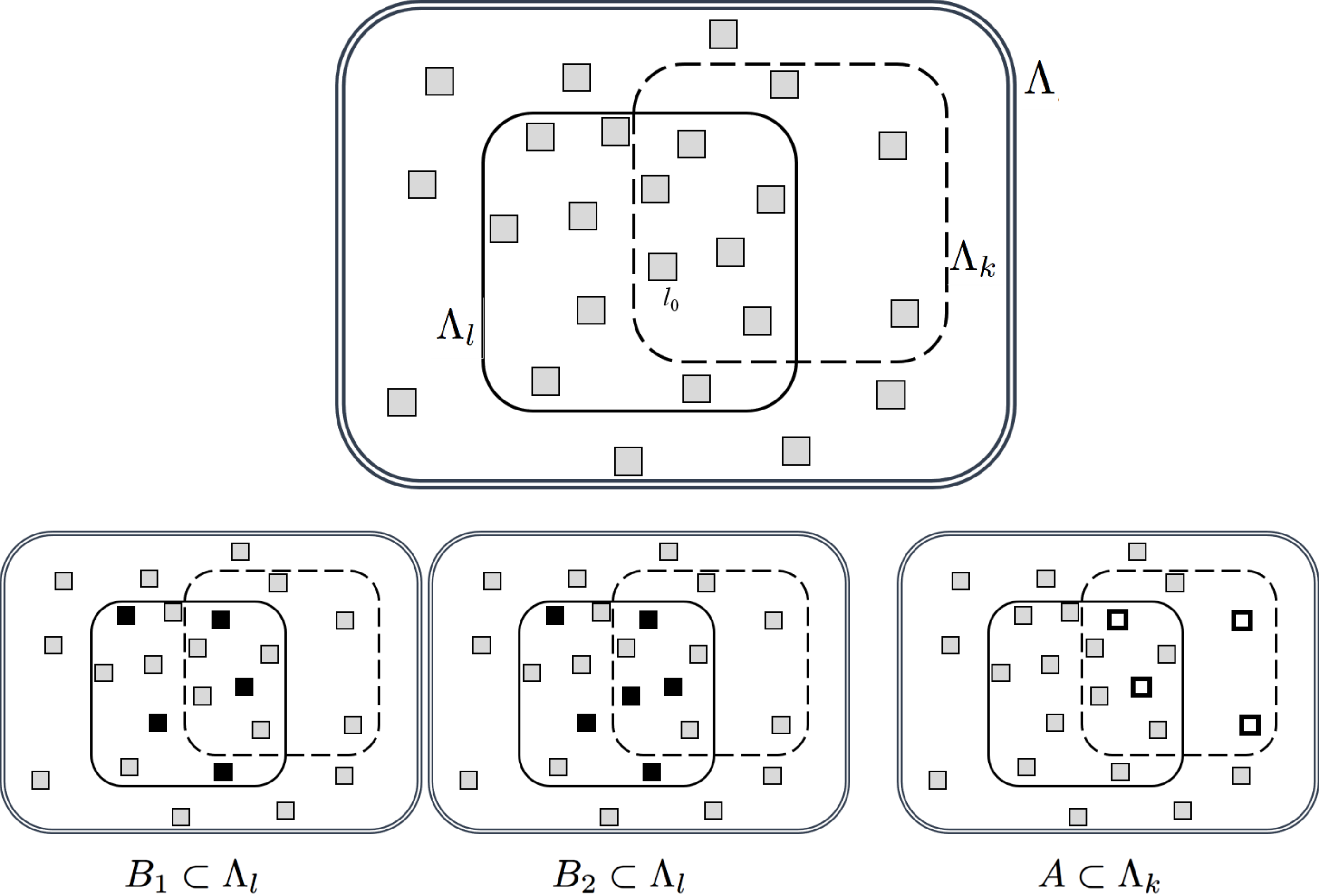}
	\caption{Examples of consistent and inconsistent local patterns: local pattern $B_1$ of link $l$ (left) is compatible with local pattern $A$ of link $k$ (right), while local pattern $B_2$ (middle) is not. Using the binary activation variables, the conflict between $B_2$ and $A$ is enforced by  imposing the constraint  $q_l^{B_2,m}+q_k^{A,m}\le 1$. 
		Gray squares correspond to inactive links, filled black squares correspond to links active in local patterns 
		of link $l$, and white squares with black outlines correspond links active in local patterns 
		of link $k$.  
	}
	\label{fig:local_global_patterns}
\end{figure}

For the throughput calculations to hold, local patterns included in a global pattern must be consistent with each other. To enforce this constraint, we introduce a  discrete activation parameter for each local pattern, denoted by $q_l^{B,m}$, which is a {\it binary} variable that takes the value of 1 when its corresponding local pattern is active in global pattern $P_m$ and 0 otherwise. Activation is enforced by the inequality,
\begin{equation*} x_l^{B,m}\le q_l^{B,m},  \end{equation*}
and consistency is enforced by limiting the sum of \textit{incompatible} local patterns to 1, allowing at most one of them to be non-zero.
The local patterns $B\subset \Lambda_l$ and $A\subset \Lambda_k$, corresponding to activation variables  $q_l^{B,m}$ and $q_k^{A,m}$, are ``compatible" if and only if,
\[  A\cap \Lambda_l = B \cap \Lambda_k, \]
which means any active link in $B$ that happens to be in the neighborhood of link $k$ is also active in $A$, i.e., the two patterns do not impose contradictory activations on any links in the overlap of their neighborhoods.
An example of consistent local patterns is shown in Fig. \ref{fig:local_global_patterns} with neighborhoods depicted as sets and links as elements of these sets. Assuming gray squares are inactive links, local pattern $A$ in the neighborhood of link $k$ is consistent with pattern $B_1$ in the neighborhood of link $l$ but inconsistent with pattern $B_2$ in the same neighborhood, since link $l_0$ is inactive in $A$ but active in $B_2$.  

Using the binary activation parameters defined above, consistency of local allocations can be enforced by the inequality,
\begin{equation} q_l^{B,m} + q_k^{A,m} \le 1, \quad \forall A\cap \Lambda_l \neq B \cap \Lambda_k. \label{eq:consistency} \end{equation}

Thus the optimization problem of (\ref{eq:basic_formulation}) can be reformulated as: 

\begin{subequations}
	\label{eq:scalable_formulation}
	\begin{align}
	&\maximize_{[x_l^{B,m}],[q_l^{B,m}],[y_m],[r_l],d}
	\qquad d \label{eq:scalable_objective} \\
	&\subjectto  \nonumber \\
	&\quad r_l = \sum_{m\in M} \sum_{B\subset \Lambda_l:l\in B} x_l^{B,m} \gamma_l^B, \qquad \qquad \quad  l\in \Lambda  \label{eq:scalable_rate} \\
	&\quad \sum_{l\in I_i}  r_l - \sum_{k\in O_i} r_k \ge \alpha_i d, \qquad \qquad \qquad \quad   i \in \Gamma \label{eq:scalable_flow} \\
	&\quad \sum_{B\subset \Lambda_l}x_l^{B,m} \le y_m , \qquad \qquad \qquad  l\in\Lambda, m\in M \label{eq:scalable_local_resource}  \\
	&\quad \sum_{m\in M} y_m \leq 1, \label{eq:scalable_global_resource} \\
	&\quad x_l^{B,m} \le q_l^{B,m}, \qquad \qquad l\in\Lambda, B\subset \Lambda_l, m\in M \label{eq:scalable_act_enf} \\
	&\quad q_l^{B,m} + \sum_{\substack{A\subset \Lambda_k \\B \cap \Lambda_k \ne A \cap \Lambda_l}} q_k^{A,m} \le 1,  \label{eq:scalable_consistency} \\
	& \qquad \qquad \qquad \qquad \qquad\,  l,k\in\Lambda, B\subset \Lambda_l,  m\in M \nonumber \\
	&\quad q_l^{B,m} \in \{0,1\}, \qquad \qquad l\in\Lambda, B\subset \Lambda_l,  m\in M \nonumber \\
	&\quad x_l^{B,m}\ge 0. \qquad \qquad \qquad l\in\Lambda, B\subset \Lambda_l,  m\in M \nonumber
	\end{align}  
\end{subequations}
In this formulation, $r_l$ is the data rate of link $l$ and (\ref{eq:scalable_flow}) is the set of flow constraints that guarantees a minimum downlink throughput of $\alpha_i d$ to non-gateway node $i$. Note that many of the consistency  constraints of (\ref{eq:consistency}) have been bundled into a single inequality in (\ref{eq:scalable_consistency}); this is possible because only one local pattern is active for each link in each slot, meaning these constraints can be compounded for local patterns of a single link.

The allocation that emerges from solving this problem is constructed as follows. The $m$-th global activation pattern is obtained by
\[ P_m=\bigcup_{l\in\Lambda,B\subset\Lambda_l : q_l^{B,m}=1}B,
\]
and is alloted a time slot of length $y_m$ normalized to the total frame. 
Using Theorem \ref{theorem:caratheodory}, it can be shown that any solution to (\ref{eq:basic_formulation}) has an $(N+1)$-sparse equivalent that is further equivalent, in terms of global patterns and allocations, to a solution of (\ref{eq:scalable_formulation}). We omit the proof and refer the reader to \cite{zhou20171000} for the proof technique.

The objective function in (\ref{eq:scalable_formulation}) can also be modified to incorporate delay penalization, by rewriting it as
\begin{align*}
&d-\lambda \sum_{l\in\Lambda} r_l  =\,  d-\lambda \sum_{m\in M} \sum_{l\in\Lambda} \sum_{B\subset \Lambda_l:l\in B} x_l^{B,m} \gamma_l^B.
\end{align*}
The second term is the sum of data rate on all links, and the weighting factor $\lambda$ is chosen with the same threshold as derived for the combinatorial formulation in (\ref{eq:priority}).

Both the objective and constraints of this formulation are linear, while the optimization variables are a mixture of continuous and discrete (binary) variables. Thus the exponentially growing linear program of (\ref{eq:basic_formulation}) is reduced to a polynomially scaling mixed integer linear program.

The number of variables in this formulation   grows polynomially with network size due to the fact that as  network size grows, {\it neighborhood} sizes remain the same. If the neighborhood size is no greater than $\mu$, the number of local variables $x_l^{B,m}$ will be no more than $N\times L\times 2^{\mu}$. This brings the number of variables in (\ref{eq:scalable_formulation}) to a total of fewer than $N\times L\times 2^{\mu}+N+L+1$ continuous and $N\times L\times 2^{\mu}$ integer values.

Similar to the combinatorial formulation, the above problem can also be modified to include both downlink and uplink by maintaining different link rate variables utilized for the two directions of service. The resulting allocation problem is formulated below.
\begin{subequations}
	\label{eq:scalable_formulation_ud}
	\begin{align}
	&\maximize_{[x_l^{B,m}],[q_l^{B,m}],[y_m],[r^d_l],[r^u_l],c}
	\qquad c \label{eq:scalable_objective_ud} \\
	&\subjectto \nonumber \\
	&\quad r^d_l + r^u_l = \sum_{m\in M} \sum_{B\subset \Lambda_l:l\in B} x_l^{B,m} \gamma_l^B, \qquad \, \, \, l\in \Lambda \label{eq:scalable_rate_ud} \\
	& \quad \sum_{l\in I_i}  r^\text{d}_l - \sum_{k\in O_i} r^\text{d}_k
	\ge \alpha_i c, \qquad \qquad \qquad \quad  i \in \Gamma \label{eq:scalable_flow_dl} \\
	& \quad \sum_{l\in O_i}  r^\text{u}_l - \sum_{k\in I_i} r^\text{u}_k
	\ge \beta_i c,
	\qquad \qquad \qquad \quad  i \in \Gamma \label{eq:scalable_flow_ul} \\
	&\quad \sum_{B\subset \Lambda_l}x_l^{B,m} \le y_m , \qquad \qquad \qquad  l\in\Lambda, m\in M \label{eq:scalable_local_resource_ud}  \\
	&\quad \sum_{m\in M} y_m \leq 1, \label{eq:scalable_global_resource_ud} \\
	&\quad x_l^{B,m} \le q_l^{B,m}, \qquad \qquad l\in\Lambda, B\subset \Lambda_l, m\in M \label{eq:scalable_act_enf_ud} \\
	&\quad q_l^{B,m} + \sum_{\substack{A\subset \Lambda_k \\B \cap \Lambda_k \ne A \cap \Lambda_l}} q_k^{A,m} \le 1,  \label{eq:scalable_consistency_ud} \\
	& \qquad \qquad \qquad \qquad \qquad \,   l,k\in\Lambda, B\subset \Lambda_l,  m\in M \nonumber \\
	&\quad q_l^{B,m} \in \{0,1\}, \qquad \qquad l\in\Lambda, B\subset \Lambda_l,  m\in M \nonumber \\
	&\quad x_l^{B,m} \ge 0, \qquad \qquad \quad \, \, \, \,  l\in\Lambda, B\subset \Lambda_l,  m\in M \nonumber \\
	&\quad r^d_l\ge 0, \quad  r^u_l\ge 0. \qquad \qquad \qquad \qquad \qquad l\in \Lambda \nonumber
	\end{align}  
\end{subequations}

Although the localized formulation is scalable in terms of problem size, unlike (\ref{eq:global_formulation}) (which is a linear program) this reformulation requires solving a mixed integer (binary) linear program which is inherently a combinatorial problem and NP-hard. While effective techniques have been devised for solving such problems, there are no guarantees for their computational efficiency. In the next section, some observations regarding  scalability and behavior of the two optimization frameworks of (\ref{eq:basic_formulation}) and (\ref{eq:scalable_formulation}) are discussed.

\section{Results and Discussion \label{sec:results}}

Solving the linear program formulated in (\ref{eq:basic_formulation}) is relatively straightforward using standard techniques such as the simplex method. Standard solvers such as the CVX package in Matlab or Gurobi were used to solve this problem for small networks of up to 15 links. However, due to  exponential growth of the problem size, enumerating the variables and building the inequality matrices quickly becomes impractical. The memory required for encoding a network with twenty bidirectional links ($L=40$) exceeds hundreds of Gigabytes, and solving a network larger than this threshold is practically impossible. In \cite{rasekh2015interference}, this problem is sidestepped by clustering the network around gateways and solving each cluster independently. This provides a suboptimal solution wherein heuristics must be employed to determine cluster association. 

The scalable formulation of (\ref{eq:scalable_formulation}), on the other hand, is a non-convex optimization problem due to the presence of integer variables. Approaches such as the branch-and-bound algorithm are generally effective methods for solving integer linear programs, but provide no guarantees for computational efficiency
. For moderately large problems such as the 4-gateway network of Fig. \ref{fig:topology_urban}
, solving (\ref{eq:scalable_formulation}) using standard solvers (such as the CVX package in Matlab or the Gurobi solver) takes an impractical amount of time. However, we find that in practice, due to sparsity of the interference matrix, the number of global patterns that are activated in the solution can be much smaller than the upper bound guaranteed by Theorem \ref{theorem:caratheodory}, so that the number of global time slots can be set to a value $T < N$, indexed by the truncated set $M_T = \{1,...,T\}$. Choosing an appropriate truncation level, $T$, involves a trade-off between computation time and throughput, as discussed in the next section.

\subsection {Computational efficiency and scaling}

\begin{figure*}%
	\subfigure[ ]{
		\label{fig:truncation_scaling_NYC}%
		\includegraphics[width=0.33\textwidth]{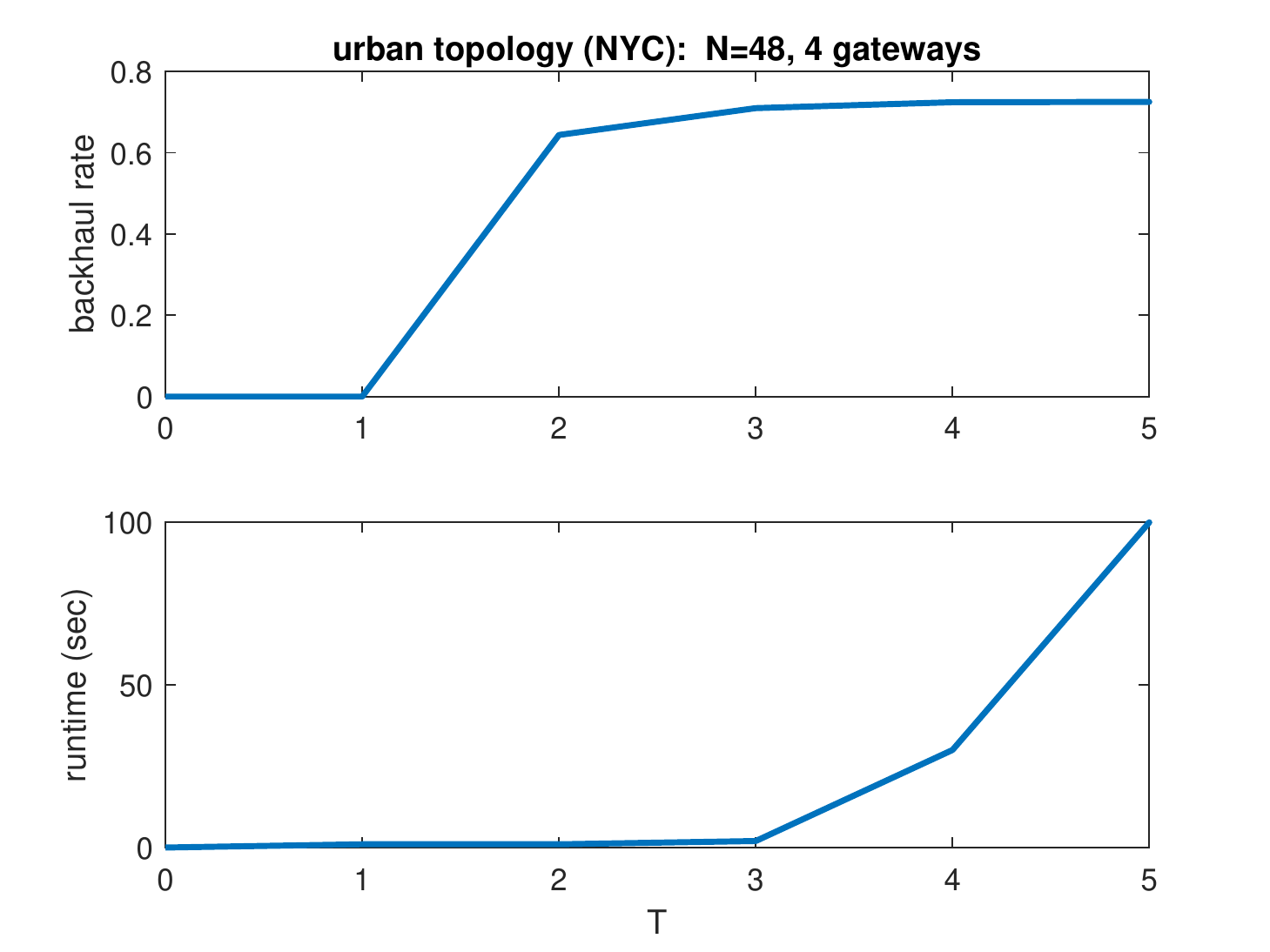}%
	}
	\hspace*{\fill}
	\subfigure[ ]{%
		\label{fig:truncation_scaling_50} %
		\includegraphics[width=0.33\textwidth]{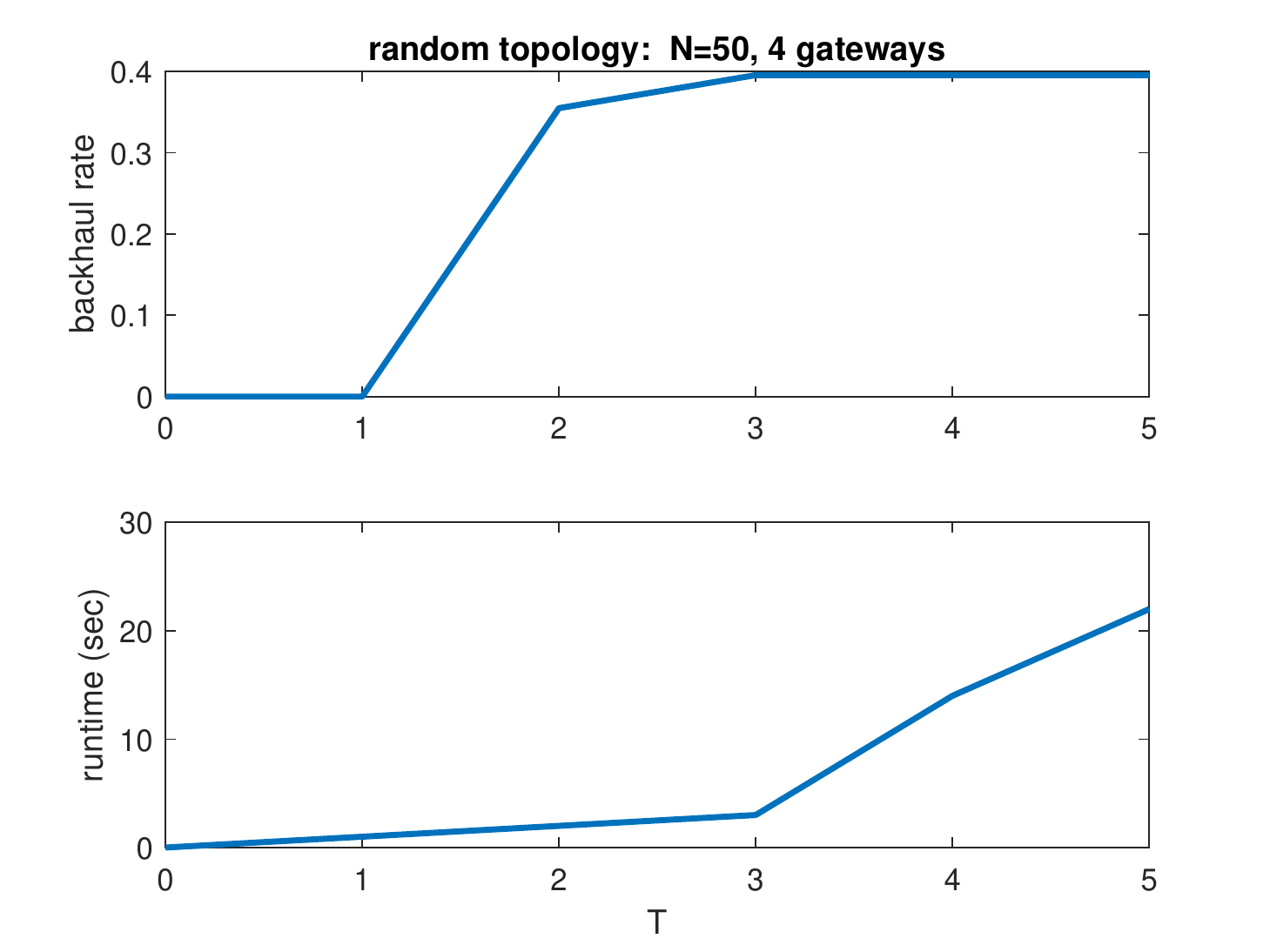}%
	}%
	\hspace*{\fill}
	\subfigure[]{
		\label{fig:truncation_scaling_100}%
		\includegraphics[width=0.33\textwidth]{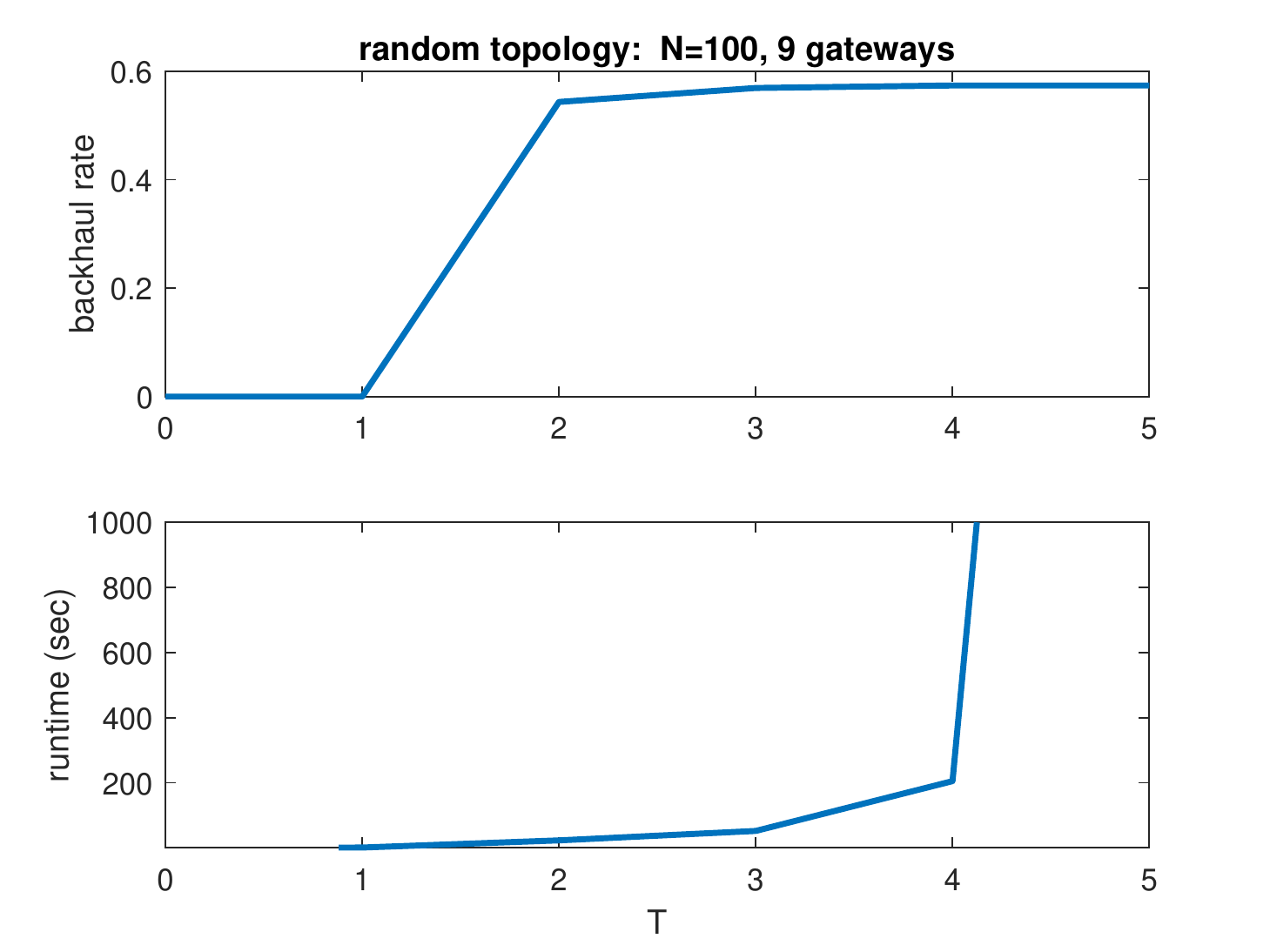}%
	}%
	\caption{Effect of truncating $M$ on obtained throughput and computation time. Nominal link rate is 3.46 (SNR=10dB). Depending on network topology, the minimum backhaul data rate delivered to every node is between 10\% and 20\% of nominal link rate.} 
	\label{fig:truncation_scaling}
\end{figure*}

Limiting the number of global patterns to a small number significantly reduces the computation time while achieving virtually all of the optimal throughput. 
It is also attractive in terms of implementation.
Fig. \ref{fig:truncation_scaling} depicts this trend, showing runtime and backhaul throughput obtained from solving (\ref{eq:scalable_formulation}) with different values for the number of global slots, $T$, in the urban network of Fig. \ref{fig:topology_urban} as well as randomly generated suburban networks of 50 and 100 nodes similar to the structure depicted in Fig. \ref{fig:topology_rural}. 
Note that the networks used in these simulations are randomly generated. Actual designed networks may have better characteristics in terms of node and gateway placement, as well as better choice of neighbor association, that prevent throughput bottlenecks and produce a more connected network. This would result in more uniform distribution of service among nodes and a higher max-min optimum throughput.

\begin{figure}
	\centering
	\includegraphics[width=.7\columnwidth]{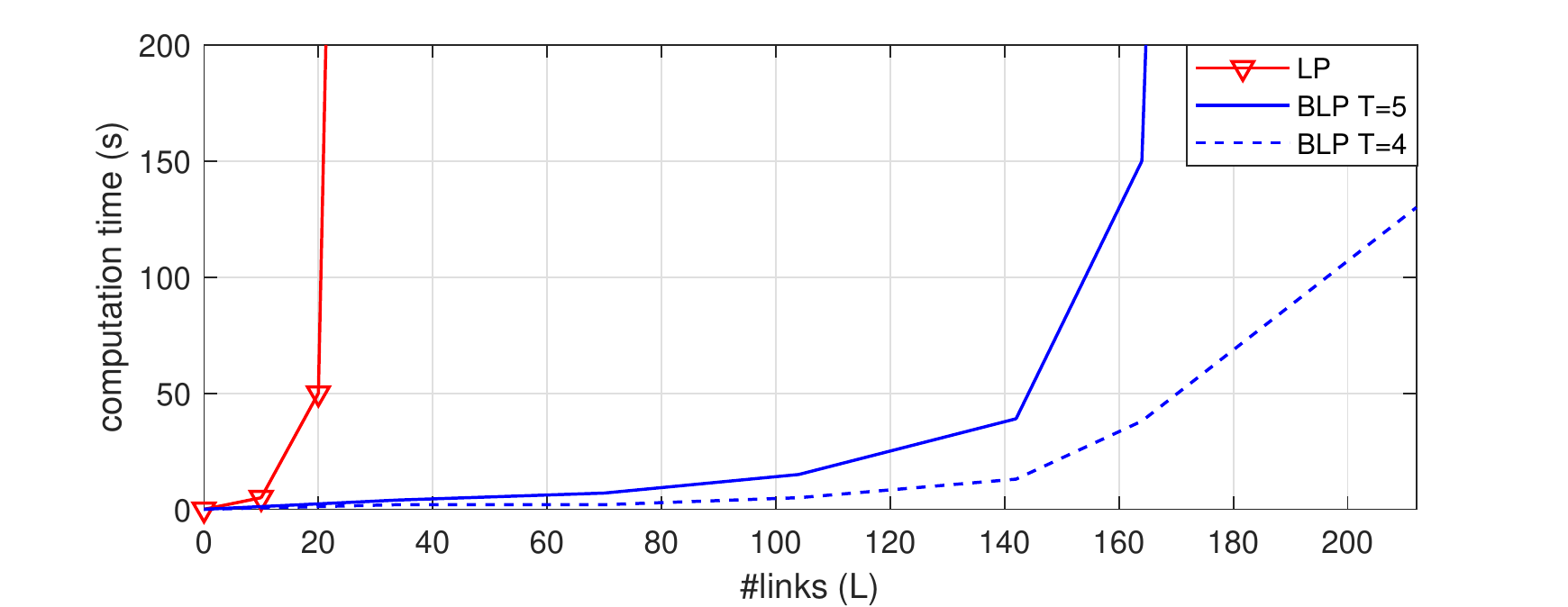}
	\caption{Computation time as a function of network size for combinatorial and truncated localized formulation. } \label{fig:scaling_with_net_size}
\end{figure}

Fig. \ref{fig:scaling_with_net_size} shows runtime as a function of network size for the original and scalable formulation (with the number of global slots, $T$, limited to 4). Networks of different sizes are generated using the suburban environment model. Formulation (\ref{eq:basic_formulation}) blows up exponentially with network size, whereas the BLP of (\ref{eq:scalable_formulation}) can be used to optimize networks of up to hundreds of nodes within a time-scale of minutes. 
In running the optimization, we found that by relaxing the rate equation of (\ref{eq:scalable_rate}) to inequality ($r_l \le \sum_{m} \sum_{B} x_l^{B,m} \gamma_l^B$), computation time decreased considerably. The results presented here were derived with this relaxation.

\subsection{Effect of residual interference}
\begin{figure}
	\centering
	\includegraphics[width=.7\columnwidth]{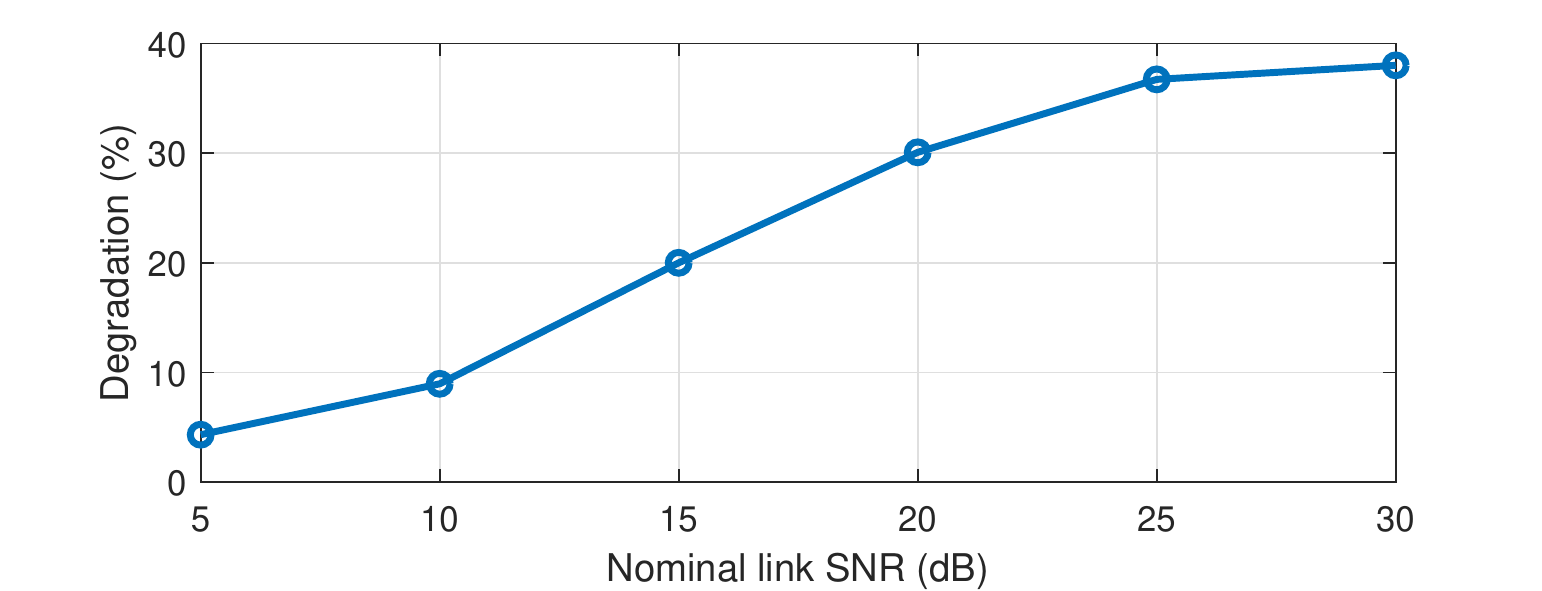}
	\caption{Degradation of throughput 
		due to suboptimal interference-agnostic scheduling, as a function of nominal link SNR (typical example; numbers derived by evaluating (\ref{eq:scalable_formulation}) on the urban network of Fig. \ref{fig:topology_urban}).} 
	\label{fig:degradation_vs_SNR}
\end{figure}

To quantify the effect of interference on backhaul capacity, we compare the throughput obtained in the two cases of (a) only modeling the half-duplex constraint (collocated TX-RX interference), and (b) including the full interference model described in Section \ref{sec:system_model}, for the urban network of Fig. \ref{fig:topology_urban}. We find that, similar to results reported in \cite{kulkarni2018many} and \cite{rasekh2015interference}, the optimal throughput capacity is the same in both cases. In fact, the half-duplex nature of transmissions requires that any highly utilized (bottleneck) link at most be activated for a fraction of the time, so that the transmitted data can be \textit{relayed} on the next link(s) on the multihop path in the remainder of the frame. This redundancy in link activation provides room for scheduling links such that no two interfering links are activated simultaneously. Thus backhaul capacity is not degraded from interference, but can only be obtained by careful scheduling of interfering links, and including interference in the model used for solving the allocation problem is crucial for obtaining such a schedule. In fact, if an interference-agnostic allocation is deployed in the network of Fig. \ref{fig:topology_urban}, the resulting max-min backhaul throughput is degraded by around 20\% when evaluated in the presence of interference. This degradation becomes more severe as link SNR increases, as depicted in Fig. \ref{fig:degradation_vs_SNR}.

\subsection{Backhaul capacity in downlink and uplink} 

The backhaul throughput provided by the network differs depending on the density of gateway nodes, number of backhaul links connected to each gateway, network structure, and directionality of antennas. For a 10:1 ratio between non-gateway and gateway nodes, approximately 20\% of the nominal link data rate can be delivered to nodes as downlink backhaul. 
Comparing with an existing network in which every base station is directly wired to the Internet, mm-wave wireless backhaul enables 10X shrinking of cell sizes by adding non-gateway base stations, resulting in significantly improved spatial reuse.
The immense capacity of densely deployed picocellular access points predicted in \cite{marzi2015interference} can thus be realized using wireless mesh backhauling, 
as long as the backhaul link data rates are high enough. 
For 10X increase in access point density, LTE cells with cell traffic in the order of hundreds of Mbps (or 1 Gbps with carrier aggregation) can be supported using wireless links with data rate of several Gbps, which is possible using the unlicensed 60 GHz band. 
On the other hand, high speed picocells that provide multi-Gbps mmwave-to-mobile access links may require tens of Gbps of backhaul throughput. This, in turn, would require backhaul links with raw data rates of the order of 100 Gbps, which could be realized using mm-wave or THz bands above 100 GHz. 

\begin{figure}
	\centering
	\includegraphics[width=.7\columnwidth]{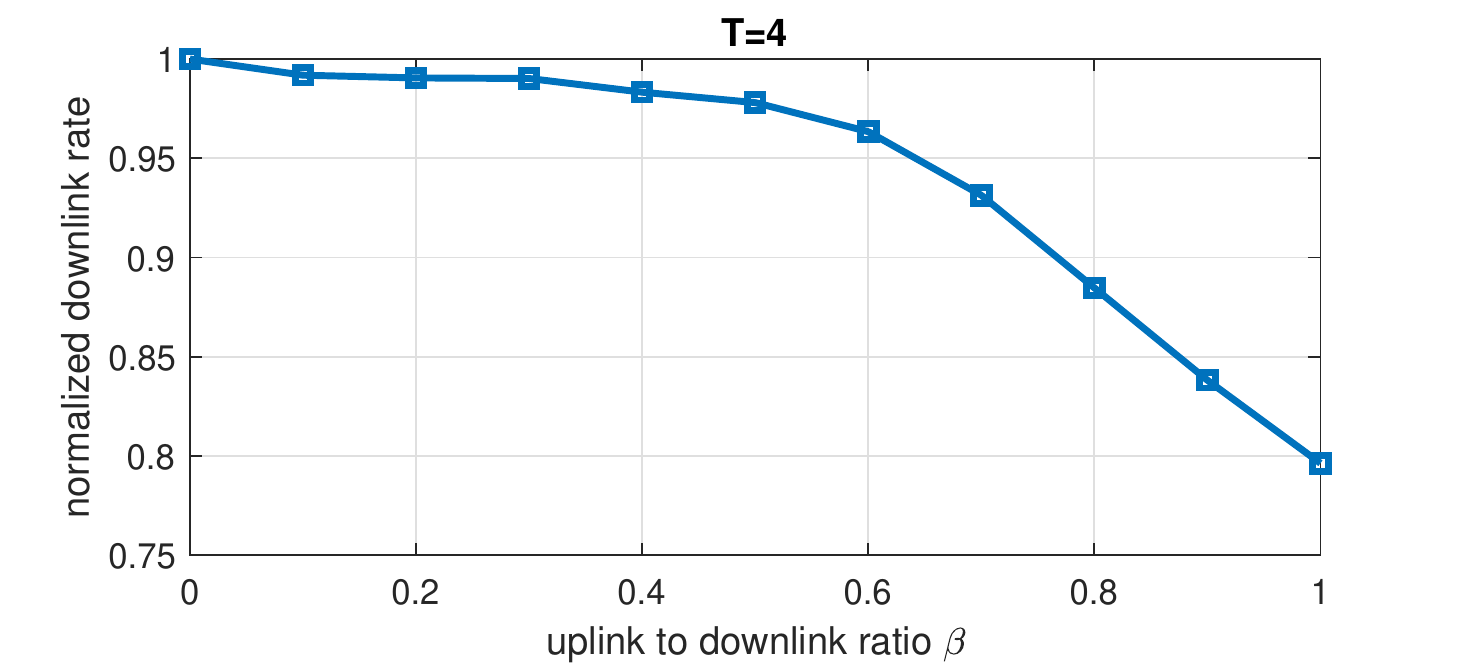}
	\caption{Degradation of downlink service rate as a function of uplink to downlink ratio.} 
	\label{fig:uplink_downlink}
\end{figure}

To quantify the joint downlink and uplink throughputs, we fix the downlink ratios $\alpha_i$ to unity and solve for different values of uniformly distributed uplink ratios $\beta_i=\beta$. We observe that an uplink ratio of up to $\beta=0.6$ can be supported with less than 5\% degradation of downlink throughput. The trade-off between downlink and uplink rate is depicted in Fig. \ref{fig:uplink_downlink} for the urban network of Fig. \ref{fig:topology_urban}. Providing equal uplink and downlink capacity results in a throughput reduction of only 20\% relative to  downlink-only support. This is expected; the link capacity that is idle because of half-duplex relaying is effectively utilized by the uplink traffic that flows in the opposite direction, without much interference on the links carrying downlink traffic.

\begin{figure}
	\centering
	\includegraphics[width=.65\columnwidth]{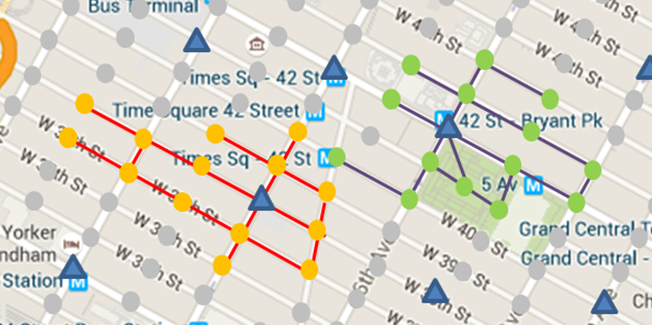}
	\caption{Example clusters formed by associating links to nearest gateway.} 
	\label{fig:NYC_clustered}
\end{figure}

\textit{Cost of clustering:} Solving the 4-gateway network of Fig. \ref{fig:topology_urban} results in a downlink rate of 21\% of nominal link rate, whereas solving for the two clusters depicted in Fig. \ref{fig:NYC_clustered} independently provides throughput of 17\%. Thus as network size grows, optimizing the network directly is preferable to clustering and provides higher throughput.

\section{Conclusions \label{sec:conclusions}}
In this paper, a multihop mm wave mesh network has been proposed for wireless backhaul of urban and suburban picocells. For joint resource scheduling and routing, 
a scalable formulation in the form of a mixed integer linear program was constructed that is able to solve moderately large networks in a time scale of minutes using a standalone PC; fast enough to adapt to slow varying traffic and topology variations. We observed that using the high speed backhaul links and angular isolation realized by mm wave antenna arrays, an order of magnitude increase in base station density can be supported. 
We also demonstrated that the formulation can be extended to jointly optimize uplink and downlink and observed that due to the redundancy caused by half duplex traffic relaying, uplink data can be routed without degradation to downlink throughput. Simulation results were reported for urban and suburban environment models, with interference models that account for building blockage and antenna patterns in the respective scenarios. 

We note that the proposed infrastructure, while designed here for the task of providing picocellular backhaul, can more generally provide
an efficient and secure backbone for a myriad of applications for ``smart cities" and metropolitan-wide internet-of-things, such as real-time traffic monitoring and control, support for vehicular communication and autonomy, surveillance, and public transportation, to name a few.  Extension of our optimization framework to
incorporate a blend of applications with different levels of delay sensitivity is an interesting topic for exploration.

Possible future directions of research also include speeding up the scheduling procedure using iterative optimization approaches or dynamic programming, possibly at the cost of slightly suboptimal solutions. Jointly optimizing the placement of nodes and neighbor associations can also improve backhaul performance.

\section{Acknowledgments}
This research was supported in part by the National Science Foundation under grants CCF-1423040, CNS-1317153, and CNS-1518812, and gifts from Facebook, Qualcomm, and Futurewei Technologies.

\bibliographystyle{IEEEtran}
\bibliography{references/refs}

\end{document}